\documentclass[12pt]{article}
\usepackage{epsfig}
\usepackage{amsmath}
\usepackage{hhline}
\usepackage{amssymb}
\usepackage{times}
\usepackage{cite}
\usepackage[]{lineno}
\usepackage[]{color}
\DeclareMathAlphabet{\mathbold}{OML}{txr}{b}{it}

\newlength{\dinwidth}
\newlength{\dinmargin}
\begin{document}  
\newcommand{\pom}{{I\!\!P}}
\newcommand{\reg}{{I\!\!R}}
\newcommand{\slowpi}{\pi_{\mathit{slow}}}
\newcommand{\fiidiii}{F_2^{D(3)}}
\newcommand{\fiidiiiarg}{\fiidiii\,(\beta,\,Q^2,\,x)}
\newcommand{\n}{1.19\pm 0.06 (stat.) \pm0.07 (syst.)}
\newcommand{\nz}{1.30\pm 0.08 (stat.)^{+0.08}_{-0.14} (syst.)}
\newcommand{\fiidiiiful}{F_2^{D(4)}\,(\beta,\,Q^2,\,x,\,t)}
\newcommand{\fiipom}{\tilde F_2^D}
\newcommand{\ALPHA}{1.10\pm0.03 (stat.) \pm0.04 (syst.)}
\newcommand{\ALPHAZ}{1.15\pm0.04 (stat.)^{+0.04}_{-0.07} (syst.)}
\newcommand{\fiipomarg}{\fiipom\,(\beta,\,Q^2)}
\newcommand{\pomflux}{f_{\pom / p}}
\newcommand{\nxpom}{1.19\pm 0.06 (stat.) \pm0.07 (syst.)}
\newcommand {\gapprox}
   {\raisebox{-0.7ex}{$\stackrel {\textstyle>}{\sim}$}}
\newcommand {\lapprox}
   {\raisebox{-0.7ex}{$\stackrel {\textstyle<}{\sim}$}}
\def\gsim{\,\lower.25ex\hbox{$\scriptstyle\sim$}\kern-1.30ex%
\raise 0.55ex\hbox{$\scriptstyle >$}\,}
\def\lsim{\,\lower.25ex\hbox{$\scriptstyle\sim$}\kern-1.30ex%
\raise 0.55ex\hbox{$\scriptstyle <$}\,}
\newcommand{\pomfluxarg}{f_{\pom / p}\,(x_\pom)}
\newcommand{\dsf}{\mbox{$F_2^{D(3)}$}}
\newcommand{\dsfva}{\mbox{$F_2^{D(3)}(\beta,Q^2,x_{I\!\!P})$}}
\newcommand{\dsfvb}{\mbox{$F_2^{D(3)}(\beta,Q^2,x)$}}
\newcommand{\dsfpom}{$F_2^{I\!\!P}$}
\newcommand{\gap}{\stackrel{>}{\sim}}
\newcommand{\lap}{\stackrel{<}{\sim}}
\newcommand{\fem}{$F_2^{em}$}
\newcommand{\tsnmp}{$\tilde{\sigma}_{NC}(e^{\mp})$}
\newcommand{\tsnm}{$\tilde{\sigma}_{NC}(e^-)$}
\newcommand{\tsnp}{$\tilde{\sigma}_{NC}(e^+)$}
\newcommand{\st}{$\star$}
\newcommand{\sst}{$\star \star$}
\newcommand{\ssst}{$\star \star \star$}
\newcommand{\sssst}{$\star \star \star \star$}
\newcommand{\tw}{\theta_W}
\newcommand{\sw}{\sin{\theta_W}}
\newcommand{\cw}{\cos{\theta_W}}
\newcommand{\sww}{\sin^2{\theta_W}}
\newcommand{\cww}{\cos^2{\theta_W}}
\newcommand{\trm}{m_{\perp}}
\newcommand{\trp}{p_{\perp}}
\newcommand{\trmm}{m_{\perp}^2}
\newcommand{\trpp}{p_{\perp}^2}
\newcommand{\alp}{\alpha_s}

\newcommand{\alps}{\alpha_s}
\newcommand{\sqrts}{$\sqrt{s}$}
\newcommand{\LO}{$O(\alpha_s^0)$}
\newcommand{\Oa}{$O(\alpha_s)$}
\newcommand{\Oaa}{$O(\alpha_s^2)$}
\newcommand{\PT}{p_{\perp}}
\newcommand{\JPSI}{J/\psi}
\newcommand{\sh}{\hat{s}}
\newcommand{\uh}{\hat{u}}
\newcommand{\MP}{m_{J/\psi}}
\newcommand{\PO}{I\!\!P}
\newcommand{\xbj}{x}
\newcommand{\xpom}{x_{\PO}}
\newcommand{\ttbs}{\char'134}
\newcommand{\xpomlo}{3\times10^{-4}}  
\newcommand{\xpomup}{0.05}  
\newcommand{\dgr}{^\circ}
\newcommand{\pbarnt}{\,\mbox{{\rm pb$^{-1}$}}}
\newcommand{\gev}{\,\mbox{GeV}}
\newcommand{\WBoson}{\mbox{$W$}}
\newcommand{\fbarn}{\,\mbox{{\rm fb}}}
\newcommand{\fbarnt}{\,\mbox{{\rm fb$^{-1}$}}}
\newcommand{\dsdx}[1]{$d\sigma\!/\!d #1\,$}
\newcommand{\eV}{\mbox{e\hspace{-0.08em}V}}
%
%
\newcommand{\qsq}{\ensuremath{Q^2} }
\newcommand{\gevsq}{\ensuremath{\mathrm{GeV}^2} }
\newcommand{\et}{\ensuremath{E_t^*} }
\newcommand{\rap}{\ensuremath{\eta^*} }
\newcommand{\gp}{\ensuremath{\gamma^*}p }
\newcommand{\dsiget}{\ensuremath{{\rm d}\sigma_{ep}/{\rm d}E_t^*} }
\newcommand{\dsigrap}{\ensuremath{{\rm d}\sigma_{ep}/{\rm d}\eta^*} }

\newcommand{\dstar}{\ensuremath{D^*}}
\newcommand{\dstarp}{\ensuremath{D^{*+}}}
\newcommand{\dstarm}{\ensuremath{D^{*-}}}
\newcommand{\dstarpm}{\ensuremath{D^{*\pm}}}
\newcommand{\zDs}{\ensuremath{z(\dstar )}}
\newcommand{\Wgp}{\ensuremath{W_{\gamma p}}}
\newcommand{\ptds}{\ensuremath{p_t(\dstar )}}
\newcommand{\etads}{\ensuremath{\eta(\dstar )}}
\newcommand{\ptj}{\ensuremath{p_t(\mbox{jet})}}
\newcommand{\ptjn}[1]{\ensuremath{p_t(\mbox{jet$_{#1}$})}}
\newcommand{\etaj}{\ensuremath{\eta(\mbox{jet})}}
\newcommand{\detadsj}{\ensuremath{\eta(\dstar )\, \mbox{-}\, \etaj}}

\def\Journal#1#2#3#4{{#1} {\bf #2} (#3) #4}
\def\NCA{\em Nuovo Cimento}
\def\NIM{\em Nucl. Instrum. Methods}
\def\NIMA{{\em Nucl. Instrum. Methods} {\bf A}}
\def\NPB{{\em Nucl. Phys.}   {\bf B}}
\def\PLB{{\em Phys. Lett.}   {\bf B}}
\def\PRL{\em Phys. Rev. Lett.}
\def\PRD{{\em Phys. Rev.}    {\bf D}}
\def\ZPC{{\em Z. Phys.}      {\bf C}}
\def\EJC{{\em Eur. Phys. J.} {\bf C}}
\def\CPC{\em Comp. Phys. Commun.}
\newcommand{\TeV}{\rm TeV}
\newcommand{\hdick}{\noalign{\hrule height1.4pt}}
\newcommand{\etbreit}{\mbox{$E_{t, {\rm Breit}}$}}
\newcommand{\etabreit}{\mbox{$\eta_{\mbox{\scriptsize Breit}}$}}
\newcommand{\rcone}{\mbox{$R_0$}}
\newcommand{\etb}{\mbox{$E_{T,B}~$}}
\newcommand{\etbx}{\mbox{$E_{T,B}$}}
\newcommand{\deltab}{\mbox{$\Delta_B~$}}
\newcommand{\deltabx}{\mbox{$\Delta_B$}}
\newcommand{\nb}{\mbox{$n_B~$}}
\newcommand{\nbx}{\mbox{$n_B$}}
\newcommand{\etjet}{\mbox{$E_{T,\rm jet}~$}}
\newcommand{\etjetx}{\mbox{$E_{T,\rm jet}$}}
\newcommand{\sph}{\mbox{${\rm Sph_B}$}~}
\newcommand{\sphx}{\mbox{${\rm Sph_B}$}}
\newcommand{\rconesq}{\mbox{$R_{0}^2$}}
\newcommand{\rrho}{\mbox{$R/\langle \rho \rangle$}}
\newcommand{\hlx}{\mbox{${\rm H_{l0}}$}}
%
\newcommand{\scaption}[1]{\caption{\protect{\footnotesize  #1}}}
\newcommand{\proc}[2]{\mbox{$ #1 \rightarrow #2 $}}
\newcommand{\average}[1]{\mbox{$ \langle #1 \rangle $}}
\newcommand{\av}[1]{\mbox{$ \langle #1 \rangle $}}
\newcommand{\W}{\mbox{$W$}}
\newcommand{\Q}{\mbox{$Q$}}
\newcommand{\xf}{\mbox{$x_F$}}  
\newcommand{\xp}{\mbox{$x_p$}}  
\newcommand{\xB}{\mbox{$x$}}  
\newcommand{\xg}{\mbox{$x_g$}}  
\newcommand{\y}{\mbox{$y$}}
\newcommand{\Qsq}{\mbox{$Q^2$}}
\newcommand{\pz}{\mbox{$P_{\!z}^{\star}$}}
\newcommand{\pzb}{\mbox{$P_{\!z}^{'}$}}
\newcommand{\pzmax}{\mbox{$P^{\star}_{\!z{\tiny \rm max}}$}}
\newcommand{\pzmaxb}{\mbox{$P^{'}_{\!z{\tiny \rm max}}$}}
\newcommand{\pt}{\mbox{$P^{\star}_{\! T}$}}
\newcommand{\ptb}{\mbox{$P^{'}_{\! T}$}}
\newcommand{\kt}{\mbox{$k_T$}}
\newcommand{\pth}{\mbox{$p_T^h$}}
\newcommand{\pte}{\mbox{$p_T^e$}}
\newcommand{\ptsq}{\mbox{$P_{\!T}^{\star 2}$}}
\newcommand{\as}{\mbox{$\alpha_s$}}
\newcommand{\ycut}{\mbox{$y_{\rm cut}$}}
\newcommand{\gx}{\mbox{$g(x,Q^2)$}}
\newcommand{\xpart}{\mbox{$x_{\rm part}$}}
\newcommand{\mrsdm}{\mbox{${\rm MRSD}^-$}}
\newcommand{\mrsdmp}{\mbox{${\rm MRSD}^{-'}$}}
\newcommand{\mrsdn}{\mbox{${\rm MRSD}^0$}}
\newcommand{\lambdams}{\mbox{$\Lambda_{\rm \overline{MS}}$}}
\newcommand{\cm}{\mbox{\rm ~cm}}
\newcommand{\GeV}{\mbox{\rm ~GeV~}}
\newcommand{\GeVx}{\mbox{\rm ~GeV}}
\newcommand{\MeV}{\mbox{\rm ~MeV}}
\newcommand{\GeVsq}{\mbox{${\rm ~GeV}^2~$}}
\newcommand{\GeVsqx}{\mbox{${\rm ~GeV}^2$}}
\newcommand{\nbinv}{\mbox{${\rm ~nb^{-1}}$}}
\newcommand{\pb}{\mbox{${\rm ~pb~}$}}
\newcommand{\pbx}{\mbox{${\rm ~pb}$}}
\newcommand{\pbinv}{\mbox{${\rm ~pb^{-1}}$}}
\newcommand{\jpsi}{\mbox{$J\!/\psi$}}
\newcommand{\sjpsi}{\mbox{\scriptsize $J\!/\psi$}}
\newcommand{\alsmz}{$\alpha_s (M_{Z^0})$}
\newcommand{\als}{$\alpha_s$}
\newcommand{\alsq}{$\alpha_s (Q)$}
\newcommand{\ep}{\mbox{$e^{\pm}p$}}
\newcommand{\mup}{\mbox{$\mu^{\pm}p$}}
\newcommand{\ee}{\mbox{$e^+e^-$}}
\newcommand{\pp}{\mbox{$p\bar{p}$}}
\newcommand{\fragfun}{\mbox{$1/\sigma_{\rm tot}\; d\sigma/dx_p$}}
\newcommand{\fx}{\mbox{$f(x,Q^2)$}}
\newcommand{\ftwo}{\mbox{$F_2$}}
\newcommand{\cms}{\mbox{cms}}
\newcommand{\Wgpsq}{\mbox{$W^2_{\! \gamma p}$}}
\newcommand{\Wgpd}{\mbox{$W^{\delta}_{\! \gamma p}$}}
\newcommand{\spsiel}{\mbox{$\sigma_{el}$}}
\newcommand{\Bf}{^\star}
\newcommand{\ich}{{\scriptscriptstyle i \in {\rm CH}}}
\newcommand{\mean}[1]{\left< #1 \right>}
\newcommand{\mf}{\mu_{\scriptscriptstyle F}}
\newcommand{\mi}{\mu_{\scriptscriptstyle I}}
\newcommand{\mr}{\mu_{\scriptscriptstyle R}}
\newcommand{\fmean}{\mean{F}}
\newcommand{\fpert}{\fmean^{\rm pert}}
\newcommand{\fpow}{\fmean^{\rm pow}}
\newcommand{\FG}[1]{\ForestGreen{#1}}
\newcommand{\stat}{{\rm stat.}}
\newcommand{\syst}{{\rm syst.}}
\newcommand{\Rp}{\mbox{$\not \hspace{-0.15cm} R_p$}}
\newcommand{\alphapom}{\alpha_{_{\rm I\!P}}}
\newcommand{\alphareg}{\alpha_{_{\rm I\!R}}}
\newcommand{\mx}{M_{_X}}
\newcommand{\my}{M_{_Y}}
\newcommand{\etamax}{\eta_{\rm max}}
\newcommand{\zpom}{z_{_{\rm \pom}}}
\newcommand{\xgam}{x_{\gamma}}
\newcommand{\zpomj}{z_{_{\rm \pom}}^{\rm jets}}
\newcommand{\xgamj}{x_{\gamma}^{\rm jets}}
\newcommand{\ol}[1]{\mbox{$\overline{#1}$}}
\newcommand{\order}[1]{\mbox{${\cal O}(#1)$}}
\newcommand{\ncs}{\mbox{$N_{\rm CS}~$}}
\newcommand{\ein}{\mbox{$E_{\rm in}~$}}
\newcommand{\eout}{\mbox{$E_{\rm out}~$}}
\newcommand{\einp}{\mbox{$E_{\rm in}'~$}}
\newcommand{\eoutp}{\mbox{$E_{\rm out}'~$}}
\newcommand{\qprime}{\mbox{$Q'~$}}
\newcommand{\qprimex}{\mbox{$Q^'$}}
\newcommand{\qprimesq}{\mbox{${Q'}^2~$}}
\newcommand{\qprimesqminx}{\mbox{${Q'}^2_{\min}$}}
\newcommand{\qprimerec}{\mbox{$Q'_{\hspace{-0.15cm} \rm rec}~$}}
\newcommand{\qprimerecx}{\mbox{$Q^'_{\hspace{-0.15cm} \rm rec}$}}
\newcommand{\qprimesqrec}{\mbox{${Q'}^2_{\hspace{-0.15cm} \rm rec}~$}}
\newcommand{\qprimesqrecx}{\mbox{${Q'}^2_{\hspace{-0.15cm} \rm rec}$}}
\newcommand{\xprime}{\mbox{$x'~$}}
\newcommand{\sprime}{\mbox{$s'~$}}
\newcommand{\qprimesqx}{\mbox{${Q'}^2$}}
\newcommand{\xprimex}{\mbox{$x'$}}
\newcommand{\sprimex}{\mbox{$s'$}}
\newcommand{\xprimeminx}{\mbox{$x'_{\min}$}}
\newcommand{\xme}{\mbox{$x'_{\rm ME}~$}}
\newcommand{\siglim}{\mbox{$\sigma_{\rm lim}~$}}
\newcommand{\siginst}{\mbox{$\sigma_{I}~$}}
\newcommand{\flim}{\mbox{$f_{\rm lim}~$}}
\newcommand{\finst}{\mbox{$f_{I}~$}}
\newcommand{\einst}{\mbox{$\varepsilon_I$}}
\newcommand{\ftwoc}{\mbox{$F_2^{c\overline{c}}$}}
\newcommand{\sigc}{\mbox{$\sigma_{c\overline{c}}$}}
\newcommand{\ccbar}{\mbox{$c\overline{c}$}}
\newcommand{\qqbar}{\mbox{$q\overline{q}$}}
\newcommand{\degr}{\mbox{$^\circ$}}
\newcommand{\kjet}{\mbox{$k_{T\rm{jet}}$}}
\newcommand{\xjet}{\mbox{$x_{\rm{jet}}$}}
\newcommand{\ejet}{\mbox{$E_{\rm{jet}}$}}
\newcommand{\thjet}{\mbox{$\theta_{\rm{jet}}$}}
\newcommand{\thlab}{\mbox{$\theta_{\rm{lab}}$}}
\newcommand{\ptjet}{\mbox{$p_{T\rm{jet}}$}}
\newcommand{\xb}{\mbox{$x~$}}  
\newcommand{\mz}{\mbox{$m_Z~$}}
\newcommand{\sigtotal}{\mbox{$\sigma_{\rm tot}~$}}
\newcommand{\sigtot}{\mbox{$\sigma_{\rm tot}^{\gamma^* p}$}}
\newcommand{\sigt}{\mbox{$\sigma_T~$}}
\newcommand{\sigl}{\mbox{$\sigma_L~$}}
\newcommand{\xe}{\mbox{$x_E~$}}  %
\newcommand{\knot}{\mbox{$K^0~$}}
\newcommand{\ls}{\mbox{$\lambda_s$}}
\newcommand{\mt}{\mbox{$m_T~$}}

\newcommand{\ethad}{\mbox{$E_T^{\rm had}~$}}
\newcommand{\etpar}{\mbox{$E_T^{\rm par}~$}}
\newcommand{\ptintr}{\mbox{$p_T^{\rm intr}~$}}
\newcommand{\ptfrag}{\mbox{$p_T^{\rm frag}~$}}
\newcommand{\ptrad}{\mbox{$p_T^{\rm rad}~$}}
\newcommand{\ptmax}{\mbox{$p_T^{\rm max}~$}}
\newcommand{\pmax}{\mbox{$p^{\rm max}~$}}
\newcommand{\nmax}{\mbox{$n_{\rm max}~$}}
\newcommand{\emax}{\mbox{$E^{\rm max}~$}}
\newcommand{\pzmin}{\mbox{$p_z^{\rm min}~$}}
\newcommand{\aw}{\mbox{$\alpha_w~$}}
\newcommand{\bQsq}{\ensuremath{\pmb{Q^2}}}

\newcommand{\textchanged}{}

\begin{titlepage}

\noindent
\begin{flushleft}
{\tt DESY 16-050    \hfill    ISSN 0418-9833} \\
{\tt March 2016}                  \\
\end{flushleft}

\noindent

\vspace{2cm}
\begin{center}
\begin{Large}

{\bf Search for QCD Instanton-Induced Processes at HERA \\
  in the High-$\bQsq$ Domain  \\}

\vspace{2cm}

H1 Collaboration

\end{Large}
\end{center}

\vspace{2cm}

\begin{abstract}
 Signals of QCD instanton-induced processes are searched for in neutral current deep-inelastic scattering 
 at the electron-proton collider HERA in the kinematic region defined by 
 the Bjorken-scaling variable $x > 10^{-3}$, 
 the inelasticity $0.2< y < 0.7$ and the photon virtuality $150 < Q^2 < 15000$ GeV$^2$. 
 The search is performed using H1 data corresponding to an integrated luminosity of ~$351$ pb$^{-1}$.
 No evidence for the production of QCD instanton-induced events is observed.
 Upper limits on the cross section for instanton-induced processes 
 between $1.5$~pb and $6$~pb, at $95\%$~ confidence level,
 are obtained depending on the kinematic domain in which instantons could be produced.
 Compared to earlier publications, the limits are improved by an order of magnitude
 and for the first time are challenging predictions. 
\end{abstract}

\vspace{1.5cm}

\begin{center}
Accepted by \EJC
\end{center}

\end{titlepage}

%
%
%
\begin{flushleft}

V.~Andreev$^{20}$,             
A.~Baghdasaryan$^{32}$,        
K.~Begzsuren$^{29}$,           
A.~Belousov$^{20}$,            
A.~Bolz$^{12}$,                
V.~Boudry$^{23}$,              
G.~Brandt$^{42}$,              
V.~Brisson$^{22}$,             
D.~Britzger$^{10}$,            
A.~Buniatyan$^{2}$
% ...


\bigskip{\it
 $ ^{1}$ I. Physikalisches Institut der RWTH, Aachen, Germany \\
 $ ^{2}$ School of Physics and Astronomy, University of Birmingham,
          Birmingham, UK$^{ b}$ \\
% ...
$ ^{10}$ DESY, Hamburg, Germany \\
% ...
$ ^{12}$ Physikalisches Institut, Universit\"at Heidelberg,
          Heidelberg, Germany$^{ a}$ \\
% ...
$ ^{20}$ Lebedev Physical Institute, Moscow, Russia \\
% ...
$ ^{22}$ LAL, Universit\'e Paris-Sud, CNRS/IN2P3, Orsay, France \\
 $ ^{23}$ LLR, Ecole Polytechnique, CNRS/IN2P3, Palaiseau, France \\
% ...
$ ^{29}$ Institute of Physics and Technology of the Mongolian
          Academy of Sciences, Ulaanbaatar, Mongolia \\
% ...
$ ^{32}$ Yerevan Physics Institute, Yerevan, Armenia \\
% ...
$ ^{42}$ Now at II. Physikalisches Institut, Universit\"at G\"ottingen,
          G\"ottingen, Germany \\
% ...
$ ^a$ Supported by the Bundesministerium f\"ur Bildung und Forschung, FRG,
      under contract numbers 05H09GUF, 05H09VHC, 05H09VHF,  05H16PEA \\
 $ ^b$ Supported by the UK Science and Technology Facilities Council,
      and formerly by the UK Particle Physics and
      Astronomy Research Council \\
% ...
}
\end{flushleft}
\newpage

\section{Introduction}

The Standard Model of particle physics contains  certain anomalous processes induced by instantons
which violate the conservation of baryon and lepton number ($B + L$) 
in the case of electroweak interactions 
and chirality in the case of strong interactions \cite{inst:belavin,inst:thooft}.   
In  quantum chromodynamics (QCD), the  theory of  strong interactions, 
instantons are non-perturbative fluctuations of the gluon field. 
They can be  interpreted as tunnelling transitions between 
topologically different vacua.
Deep-inelastic scattering (DIS) offers a unique 
opportunity~\cite{inst:vladimir} to discover a class of hard processes
induced by  QCD instantons. 
The corresponding cross section will be referred to as the instanton cross section. It is calculable
within  ``instanton-perturbation theory'' 	
and is expected to be sizable \cite{inst:balitsky1,inst:moch,inst:rs4,inst:rs-lat}.
Moreover, the instanton-induced final state exhibits a characteristic signature
\cite{inst:vladimir,inst:greenshaw,inst:mcws99,inst:schremppzoom,inst:qcdins}.
Detailed reviews are given elsewhere~\cite{inst:ringberg00,inst:ringberg01}. 
The theory overview given here follows closely the one in the previous H1
publication~\cite{Adloff:2002ph}.

An experimental observation of  instanton-induced processes 
would constitute a discovery of a basic and yet novel non-perturbative
QCD effect at high energies.
The theory and phenomenology for the production of instanton-induced processes
at HERA in neutral current (NC) electron\footnote{The term ``electron'' is used in the following 
to refer to both electron and positron.}-proton collisions  has been worked out by
Ringwald and Schrempp \cite{inst:vladimir,inst:moch,inst:rs4,inst:rs-lat,inst:greenshaw,inst:mcws99}.
The size of the predicted cross section is large enough     
to make an experimental observation possible.
The expected signal rate is, however, still small compared to that from the
standard NC DIS (sDIS) process. 
The suppression of the sDIS background
is therefore the key issue.
QCD instanton-induced processes can be discriminated from sDIS by their 
characteristic hadronic final state signature, 
consisting of a large number of hadrons at high transverse energy
emerging from a ``fire-ball''-like topology in the instanton rest system 
\cite{inst:vladimir,inst:greenshaw,inst:mcws99}.
Discriminating  observables, derived from simulation studies, are exploited to
identify a phase space region where a difference between
data and  sDIS expectations would indicate a 
contribution from instanton-induced processes.

Upper cross section limits on instanton-induced processes have been reported
by the H1 \cite{Adloff:2002ph} and ZEUS \cite{Chekanov:2003ww} collaborations.
This analysis is a continuation of the previous H1 search for QCD instanton-induced events
using  a seventeen times larger data sample. The search is carried out at significantly higher  
virtualities of the exchanged photons as suggested by theoretical 
considerations~\cite{inst:schremppzoom}.

\section{Phenomenology of QCD Instanton-Induced Processes in NC DIS}
\label{sec:qcdtheo}
\begin{figure}[h]
   \centering
\hspace{2.0cm}
\begin{tabular}{ll}
\begin{tabular}{l}
\mbox{ 
 \epsfig{file=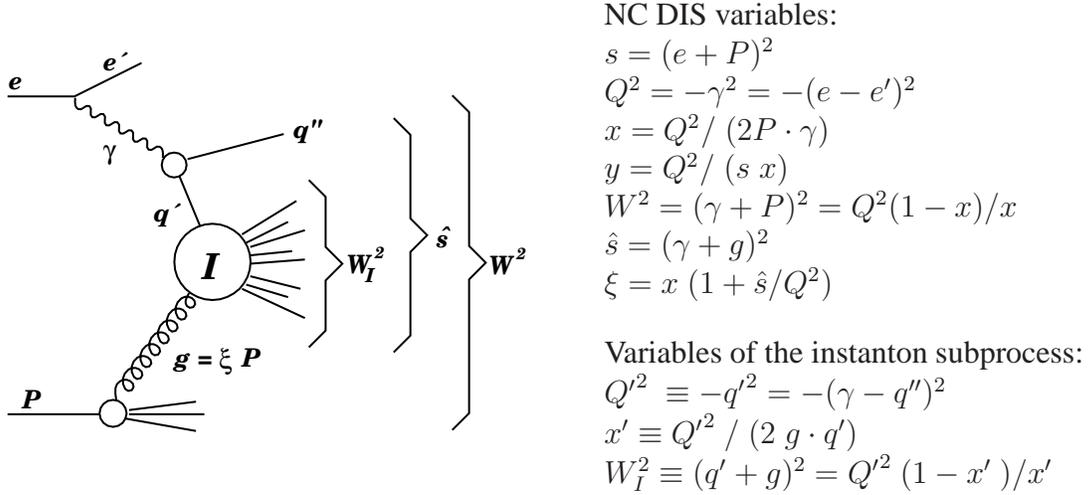  ,width=7.cm}
}
\end{tabular}
&
\begin{tabular}{l}
 NC DIS variables: \\
   $s=(e+P)^2$\\
   $\Qsq = - \gamma^2 = -(e-e')^2$ \\
   $ x = \Qsq / \; (2 P \cdot \gamma) $ \\
   $ y = \Qsq / \; (s \; x) $ \\
   $ W^2 =(\gamma+P)^2 = \Qsq (1 - x)/x$ \\
   $ \hat{s} = (\gamma+g)^2$ \\
   $ \xi = x \;(1+\hat{s}/Q^2)$ \\ \\
 \vspace{-0.6cm} \\
 Variables of the instanton subprocess: \\
 $\qprimesq  \equiv - {q'}^2 = - (\gamma-q'')^2 $ \\
 $x' \equiv \qprimesq / \;(2 \; g \cdot q' ) $ \\
 $W_I^2  \equiv (q'+g)^2 = \qprimesq ( 1 - \xprime )/ \xprime $\\
\end{tabular}
\end{tabular}
\vspace{0.3cm}
   \caption[$4$-momenta and kinematic variables of QCD instanton-induced process
     in deep-inelastic scattering]
     {Kinematic variables of the dominant instanton-induced process in DIS. 
      The virtual photon 
      ( $\gamma=e-e'$, virtuality $Q^2$), emitted by the incoming electron $e$,
      fuses with a gluon ($g$) radiated from the proton
      ($P$).
      The gluon carries a fraction $\xi$ of the longitudinal
      proton momentum. The virtual quark $(q')$ is viewed as entering the 
      instanton subprocess and the outgoing quark $q''$ from the 
      photon splitting process is viewed as the current quark. 
      The invariant mass of the quark gluon ($q'g$) system is $W_I$,
      $W$ denotes the invariant mass of the total hadronic system 
      (the $\gamma P$ system) and
      $\hat{s}$ refers to the invariant mass squared of the $\gamma g$ system.}
   \label{kin-var}
\end{figure}

Instanton  processes  
predominantly occur in  photon gluon ($\gamma g$) 
fusion processes as sketched in figure~\ref{kin-var}. 
The characteristic instanton event signatures result from the following basic 
chirality violating reaction:
\begin{equation}
\gamma^* + g \stackrel{(I)}{\rightarrow} \sum_{q=d,u,s,...} (q_R + \bar{q}_R) + \, n_g \, g,
\; \; \;  ( I \rightarrow \bar{I}, R \rightarrow L),
\end{equation}
where $g$, $q_R$ ($\bar{q}_R$) denotes gluons, 
right-handed quarks (anti-quarks),
and $n_g$ is the number of gluons produced.
The chirality violation\footnote{$\Delta \rm chirality = 2 \, n_f$, where
$\Delta \rm chirality =   $ \# $(q_{R} +\bar{q}_{R})-$ \# $(q_{L} +\bar{q}_{L})$,
 and $n_f$ is the number of quark flavours.}
is induced for each flavour, in accord with the 
corresponding axial anomaly \cite{inst:thooft}. In consequence, in every instanton event,
quark anti-quark pairs of each
of the $n_f$ flavours 
occur precisely once.
%
%
Right-handed quarks are produced in instanton-induced processes $(I)$,
left-handed quarks are produced in anti-instanton $(\bar{I}$) processes. 
The final state induced by instantons or anti-instantons
can  be distinguished only by the chirality of the quarks.
Experimental signatures sensitive to instanton-induced chirality violation
are, however, not exploited in this analysis. Both 
instanton  and  anti-instanton processes enter likewise in the calculation 
of the total cross section.

%

In photon-gluon fusion processes,
a photon splits into a quark anti-quark pair
in the background of an instanton or an anti-instanton field, as shown in figure~\ref{kin-var} .
The so-called instanton subprocess $q' + g  \stackrel{(I,\bar{I})}{\rightarrow} X$
is induced by
the quark or the anti-quark fusing with a gluon $g$ from the proton. 
The  partonic system $X$  contains $2 \, n_f $  quarks and anti-quarks, 
where one of the quarks (anti-quarks) acts as the current quark ($q''$).
In addition, an average number of 
$\langle n_g \rangle  \sim {\cal O}(1/\alpha_s)  \sim  3$ gluons is
emitted in the instanton subprocess.

The quarks and gluons emerging from the instanton subprocess are
distributed isotropically in the instanton rest system defined by
$\vec{q'} + \vec{g} = 0$.
Therefore one expects to find a
pseudo-rapidity\footnote{The pseudo-rapidity of a 
particle is defined as $\eta \equiv - \ln  \tan ( \theta / 2) $, where $\theta$ is 
the polar angle with respect to the proton direction defining the
$+z$-axis.}  ($\eta$)
region with a width of typically $2$ units in  $\eta$,
densely populated with particles of relatively high transverse
momentum and isotropically distributed in azimuth, measured in the
instanton rest frame. The large number of partons emitted in the instanton process leads
to a high multiplicity of charged and neutral particles. 
Besides this band in pseudo-rapidity, 
the hadronic final state also contains a current jet emerging from
the outgoing current quark $q''$.

The instanton production cross section at HERA,
$\sigma^{(I)}_{\rm HERA}$, is determined by 
the cross section of the instanton subprocess $q' + g  \stackrel{(I,\bar{I})}{\rightarrow} X$.
The subprocess cross section  is calculable in instanton perturbation theory.
It involves the distributions of 
the size $\rho$ of instantons and of the
distance $R$ between  them.
By confronting instanton perturbation theory with non-perturbative
lattice simulations of the QCD vacuum, limits on the validity of instanton
perturbation theory have been derived~\cite{inst:rs4,inst:rs-lat,inst:schremppzoom}.
The  perturbative and lattice calculations agree
for $\rho\lesssim 0.35 $~fm and $R/\rho\gtrsim 1.05$. At larger $\rho$ or smaller
 $R/\rho$, the instanton perturbative cross section grows, whereas  the lattice calculations suggest that the
 cross section is limited.
There is a relation between the variables $\qprime$ and $\xprime$
in momentum space and the  spatial variables $\rho$ and $R/\rho$.
Large $\qprime$  and $\xprime$ values correspond to small $\rho$ 
and large $R/\rho$, respectively.
The aforementioned limits can be translated 
into regions  of the kinematical variables \xprime and \qprimesqx, in which
the perturbative calculations are expected to be valid, 
$\qprimesqx\ge\qprimesqminx \simeq (30.8\times\Lambda^{n_f}_{\overline {MS}})^2$ 
and $\xprimex \ge\xprimeminx \simeq 0.35$~\cite{inst:qcdins}.
Here $\Lambda^{n_f}_{\overline {MS}}$ is the QCD scale
in the $\overline {MS}$ scheme for $n_f$ flavours.
In order to assure the dominance of planar diagrams
the additional restriction $Q^2 \ge \qprimesqminx$ is
recommended~\cite{inst:moch,inst:schremppzoom,inst:qcdins}.
The cross section depends  significantly 
on the strong coupling\footnote{The qualitative behaviour for the instanton cross section is
$\sigma_{q'g}^{(I)}  \sim 
{\left [\frac{2 \pi}{\alpha_{s}} \right ]}^{12} 
e^{- \frac{4 \pi}{\alpha_{s}}}$,
where $\alpha_s$ is the strong coupling.}
$\alpha_s$, or more precisely on $\Lambda^{n_f}_{\overline {MS}}$, 
but depends only weakly on the choice of the renormalisation scale.

 The calculation of the instanton production cross section in instanton perturbation theory
[5-7] is valid in the dilute instanton-gas approximation  for approximately massless
flavours, i.e. $n_f=3$, in the HERA kinematic domain. The contribution of heavy flavours is
expected to be (exponentially) suppressed \cite{Schrempp:private,Andrei:1978xg}. 
Thus calculations of the instanton production cross section 
using  the QCDINS Monte Carlo generator \cite{inst:qcdins}  are performed 
for $n_f=3$ massless flavours. It was checked that the predicted final state signature 
does not change significantly when heavy flavours are included in the simulation.

The analysis is performed in the kinematic region defined by 
$ 0.2 < y < 0.7$ and $150 < Q^2< 15000~{\rm GeV}^2$.
In this  kinematic region, and additionally requiring $\qprimesq  > 113$~\GeVsqx~and $\xprime > 0.35$,  
the cross section predicted by  QCDINS
is $\sigma^{(I)}_{\rm HERA} = 10\pm 3 \; {\rm pb}$,
using  the QCD scale
$\Lambda_{\overline{MS}}^{(3)}= 339\pm 17 \; {\rm MeV}$ \cite{Aoki:2013ldr}.
The quoted uncertainty of the instanton cross section  $\sigma^{(I)}_{\rm HERA}$
is obtained by varying the QCD scale by one standard deviation. 

The fiducial region  in $\qprimesqx$ and $\xprime$ of the validity of instanton perturbation theory
was derived from ${\rm n}_f=0$ lattice simulations, since ${\rm n}_f=3$ was not  available
for this purpose.
The perturbative instanton calculation  is made 
in the ``dilute instanton gas'' approximation, where the average distance between instantons
should  be large  compared to the instanton size. This approximation is 
valid for $\xprime\!\rightarrow\!1$, whereas the boundary $\xprime = 0.35$
corresponds to a configuration  where the distance $R$ is similar to the instanton size $\rho$.
A further simplifying assumption is made by choosing a simple form of the fiducial region
with fixed  $\qprimesqminx$ and $\xprimeminx$, whereas  $\qprimesqminx$  could be 
varied as a function of $\xprimeminx$.
In summary, the kinematic region in $\qprimesqx$ and  $\xprime$, where instanton 
perturbation theory is reliable, is, for the reasons given above,
not very well defined. Thus, the theoretical uncertainty of the instanton cross section
is difficult to define and could be larger than the already significant uncertainty
due to the uncertainty of the QCD scale $\Lambda_{\overline{MS}}^{(3)}$ alone.
On the other hand, given that the predicted cross section is large,
dedicated searches for instanton-induced processes at HERA are well motivated.

\section{Experimental Method}
\label{sec:det}

\subsection{The H1 Detector}

A detailed description of the H1 detector can be found
elsewhere~\cite{h1detector,h1tracker,h1lar,spacal}.
The origin of the H1 coordinate system is given by the nominal
$ep$ interaction point at $z=0$. The direction of the proton beam defines
the positive $z$--axis (forward direction)
and the  polar angle $\theta$ and transverse momentum $P_T$ of every particle 
is defined with respect to this axis.  
The azimuthal angle $\phi$ defines the particle direction in the transverse plane. 
The detector components most
relevant to this analysis are the Liquid Argon (LAr) calorimeter,
which measures the positions and energies of particles over the range
$4^\circ<\theta<154^\circ$ with full azimuthal coverage, the inner tracking detectors, which
measure the angles and momenta of charged particles over the range
$7^\circ<\theta<165^\circ$, and a lead-fibre calorimeter (SpaCal)
covering the range $153^\circ<\theta<174^\circ$.

The LAr calorimeter consists of an electromagnetic section with lead
absorbers and a had\-ro\-nic  section with steel absorbers.
The electromagnetic and the hadronic sections
are highly segmented in the transverse and the longitudinal
directions. Electromagnetic shower energies are measured with a
resolution of $\delta E/E \simeq 0.11/\sqrt{E/{\rm GeV}} \oplus 0.01$
and hadronic energies with $\delta E/E \simeq 0.50/\sqrt{E/{\rm GeV}}
\oplus 0.03$ as
determined using electron and pion test beam
measurements~\cite{Andrieu:1993tz,Andrieu:1994yn}.

In the central region, $15^{\circ}<\theta<165^{\circ}$, the central
tracking detector (CTD) measures the trajectories of charged particles
in two cylindrical drift chambers immersed in a uniform $1.16\,{\rm T}$
solenoidal magnetic field. In addition, the CTD  contains a  drift
chamber (COZ) to improve the $z$-coordinate
reconstruction and a multi-wire proportional
chamber at inner radii (CIP) mainly used for
triggering~\cite{Becker:2007ms}. The CTD measures charged particles
with a transverse momentum resolution of $\delta(p_T)/p_T\simeq
0.002 \, p_T/{\rm GeV} \oplus 0.015$. The forward
tracking detector (FTD) is used to supplement track reconstruction in
the region $7^{\circ}<\theta<30^{\circ}$~\cite{Laycock:2012xg}. It improves the hadronic
final state reconstruction of forward going low transverse momentum particles.
The CTD tracks are linked to hits in the vertex detector, the central
silicon tracker (CST)~\cite{h1cst,h1cst2}, to provide 
precise spatial track reconstruction.
 

In the backward region the SpaCal provides an energy measurement for
hadronic particles, and has a hadronic energy resolution of $\delta
E/E \simeq 0.70/\sqrt{E/{\rm GeV}}\oplus 0.01$ and a resolution for
electromagnetic energy depositions of $\delta E/E \simeq
0.07/\sqrt{E/{\rm GeV}}\oplus 0.01$ measured using test beam
data~\cite{spacal_res}. 

The $ep$ luminosity is determined by measuring the event rate
for the Bethe-Heitler process  $ep \rightarrow
ep\gamma$, where the photon is detected in the photon tagger located at
$z=-103\,{\rm m}$. The overall
normalisation is determined using a precision measurement of the QED
Compton process~\cite{compton-lumi} with the electron and the photon 
detected in the SpaCal.

\subsection{Data Samples}
\label{sec:datasets}

High $Q^2$ neutral current DIS events are
triggered mainly using information from the LAr calorimeter.  The
calorimeter has a finely segmented pointing geometry allowing the
trigger to select localised energy deposits in the electromagnetic
section of the calorimeter pointing to the nominal interaction vertex.
For electrons with energies above $11$~GeV the trigger efficiency 
 is determined to be close to $100\%$ \cite{h1hiq2jets}.

This analysis is performed using the full $e^{\pm}p$ collision data set taken
in the years 2003-2007 by the H1 experiment.
The data were recorded with a lepton beam of energy $27.6$~GeV and a proton beam of energy 
$920$~GeV, corresponding to a centre-of-mass energy $\sqrt{s}=319$~GeV.
The total integrated luminosity of the analysed data is $351$~pb$^{-1}$.
\subsection{Simulation of Standard  and Instanton Processes}
\label{sec:SimMC}
Detailed simulations of the H1 detector response to hadronic final
states have been performed for two QCD models of the 
sDIS  (background) and for QCD instanton-induced scattering 
processes (signal).

The background is modelled using the RAPGAP and DJANGOH Monte Carlo programs.
The RAPGAP Monte Carlo program~\cite{rapgap} incorporates the
${\cal O} (\alpha_{s})$ QCD matrix elements and 
models higher order parton emissions
 to all orders in $\alpha_s$ using the concept of parton showers~\cite{shower} 
based on the leading-logarithm DGLAP equations~\cite{dglap}, where
QCD radiation can occur before and after the hard subprocess. 
An alternative treatment of the perturbative phase is implemented
in DJANGOH~\cite{Charchula:1994kf} which uses the Colour Dipole Model~\cite{cdm}  
with QCD matrix element corrections as implemented
in ARIADNE~\cite{ariadne}.
In both MC generators  hadronisation is modelled 
with the LUND string fragmentation~\cite{lund,Sjostrand:1993yb} 
using the ALEPH tune~\cite{Schael:2004ux}. QED radiation and electroweak effects are simulated
using the HERACLES~\cite{Kwiatkowski:1990es} program, which is interfaced 
to the RAPGAP and DJANGOH event generators.
The parton density functions of the proton are taken from the CTEQ6L set ~\cite{CTEQ6L}.

QCDINS~\cite{inst:qcdins,inst:schremppdis95} is a 
Monte Carlo package to simulate 
QCD instanton-induced scattering processes in DIS.
The hard process generator is embedded in the HERWIG~\cite{herwig}
program and is implemented as 
explained in section \ref{sec:qcdtheo}. The number of flavours is set to $n_f = 3$.
Outside the allowed region  defined  by $\qprimesqminx$ and $\xprimeminx$ 
the instanton cross section is set to zero.
The CTEQ5L~\cite{CTEQ5L} parton density functions are 
employed\footnote{In the phase space of this analysis 
the  CTEQ5L and CTEQ6L gluon density distributions are almost identical.}.
Besides the hard instanton subprocess, subleading QCD emissions
are simulated in the leading-logarithm approximation, using the coherent
branching algorithm implemented in HERWIG.
The hadronisation is performed according to the Lund string fragmentation.

The generated events are passed through a detailed GEANT3~\cite{Brun:1987ma} 
based simulation of the H1 detector and subjected
to the same reconstruction and analysis chains as are used for the data.

\subsection{Inclusive DIS Event Selection}
Neutral current DIS events  are triggered and  selected by requiring a cluster 
in the electromagnetic part of the LAr calorimeter.
The scattered electron is identified as the isolated cluster  
of highest transverse momentum. 
A minimal electron energy  of 11 GeV is required.
The remaining clusters in the calorimeters and the charged tracks are attributed to 
the hadronic final state (HFS), which is reconstructed using an energy flow algorithm without double
counting of energy \cite{hadroo1,Hellwig:2004yp,hadroo2}.
The default electromagnetic energy calibration and  alignment of 
the H1 detector~\cite{h1hiq2} as well as  the HFS calibration~\cite{h1hiq2jets,Roman} are applied.
The longitudinal momentum  balance  is required to be
within $45\,\GeV  < \sum (E - p_z) < 65\,\GeVx$,
where the sum runs over the scattered electron and
all HFS objects.
Furthermore the position of the  $z$-coordinate of the reconstructed 
event vertex must be 
within $\pm 35\,\mbox{\rm cm}$ of the nominal interaction point.

The photon virtuality $Q^2$, the Bjorken scaling variable $x$ and the inelasticity 
of the interaction $y$ 
are reconstructed from the scattered electron and  the hadronic final state particles 
using the electron-sigma method~\cite{Bassler:1997tv}. This method is the most 
precise one in the kinematic range of this analysis. 
The events are selected to cover the phase space region defined by
 $ 0.2 < y < 0.7$, $x >10^{-3}$
and $150 < Q^2< 15000~{\rm GeV}^2$.

The events passing the above cuts yield the NC DIS sample which
forms the basis of the subsequent analysis.
It consists of about $350000$ events.
The simulated events are subjected to the same reconstruction and analysis chains as
the real data.
They reproduce well the shape and the absolute
normalisation of the distributions of the 
energy and angle of the scattered electron 
as well as the kinematic variables $x$, $Q^2$ and $y$.

\subsection{Definition of the Observables and the Search Strategy}
\label{sec:obs}
The observables used to discriminate the instanton-induced contribution from that
of sDIS processes are based on the hadronic final state objects and on a selection 
of charged particles. 
Only HFS objects with  $ \eta_{\rm Lab} < 3.2$ are considered.  
Charged particles are required to have  transverse momenta with $P_T^{\rm Lab} > 0.12$ \GeV 
and polar angles with $20^o < \theta < 160^o$.
Here $\eta_{\rm Lab}$ and $ P_T^{\rm Lab}$ are measured in the laboratory frame.

In the following, all HFS objects are boosted to the hadronic
centre-of-mass frame (HCM)\footnote{The hadronic centre-of-mass  frame
is defined by
$\vec{\gamma} + \vec{P} = 0$, where $\vec{\gamma}$ and $\vec{P}$ are the $3$-momentum
of the exchanged photon and proton, respectively.}.
Jets are defined by the inclusive $k_{T}$ algorithm \cite{Ellis:1993tq} 
as implemented in FastJet \cite{Cacciari:2005hq}, 
with the massless $P_{T}$ recombination scheme
and with the distance parameter $R_{0}= 1.35 \times R_{\rm cone} $. 
A cone radius  $R_{\rm cone} = 0.5$ is used. Jets are required to have transverse energy in the HCM frame
\etjetx $ > 3 $ GeV.
Additional  requirements on the transverse energy and pseudorapidity of the jets in the laboratory 
frame are imposed,
$-1.0< \eta^{\rm Lab}_{\rm Jet} < 2.5 $ and $E_{T,\rm Jet}^{\rm Lab} > 2.5 $ GeV,
in order to
ensure that jets are contained within the acceptance of the LAr
calorimeter and are well calibrated. 
The events are selected by requiring  at least one jet with \etjetx $> 4$ GeV.
The jet with the highest transverse energy 
is used to estimate the $4$-momentum $q''$ of the current quark
(see figure~\ref{kin-var}).
\qprimesq can be reconstructed from the particles associated with the current jet 
and the photon $4$-momentum, which is obtained  using the measured 
momentum of the scattered electron.
The \qprimesq resolution is about $40 \%$.
However, the distribution of the true over the reconstructed value
exhibits large tails, since in about $35\%$ of the cases the wrong jet 
is identified as the current jet.
Due to the limited accuracy of the \qprimesq reconstruction, 
the reconstructed \qprimesqx, labelled \qprimesqrecx, cannot be used to experimentally 
limit the analysis to the kinematically allowed region $\qprimesqx\gsim\qprimesqminx$.
Details of  the \qprimesq reconstruction 
are described in \cite{inst:mcws99,inst:jgerigk,inst:koblitz}. 

The hadronic final state objects belonging to the current jet are not used 
in the definition of the following observables. 
A band in pseudo-rapidity with a width of $\pm 1.1$ units in $\eta$ 
is defined around the mean
$\bar{\eta} = \sum E_T \eta /(\sum E_T)$,
where the sum includes hadronic final state objects~\cite{birger}.
This pseudo-rapidity band is referred to as the ``instanton band''.
The number of charged particles in the instanton band \nb and the total scalar transverse energy 
of all hadronic final state objects in the instanton band \etb are measured.

An approximate instanton rest frame, where all hadronic final state objects in the instanton band
are distributed isotropically, is defined by $\vec{q'} +  \xi  \vec{P} = 0$.
The definition of $\xi$ is given in  figure~\ref{kin-var}.
A numerical value of $\xi = 0.076$ is used throughout this analysis~\cite{Adloff:2002ph}.
In the instanton rest frame the sphericity~\sph 
and the first three normalised Fox-Wolfram moments 
are calculated~\cite{Fox:1978vw,Sjostrand:1993yb}.
For spherical events \sph is close to unity,
while for pencil-like events \sph tends to zero.
Furthermore, the axes
$\vec{i}_{\rm min}$ and $\vec{i}_{\rm max}$
are found for which in the instanton rest system 
the summed projections of the $3$-momenta
of all hadronic final state objects in the instanton band are minimal or
maximal \cite{inst:greenshaw}.  The relative difference between
$E_{\rm in} = {\sum_h |\vec{p}_h \cdot \vec{i}_{\rm max} |}$ and 
$E_{\rm out}= {\sum_h |\vec{p}_h \cdot \vec{i}_{\rm min} |}$ 
is called $\deltabx = (E_{\rm in}-E_{\rm out})/E_{\rm in}$.  
This quantity is a measure of the 
transverse energy weighted azimuthal isotropy of an event.  
For isotropic events  \deltab is
small while for pencil-like events \deltab is close to unity.

The reconstruction of the variable \xprime suffers from poor resolution as
in the case of \qprimesqrecx.
Using two methods to calculate the invariant mass of the quark gluon system, $W_{I}$,
$x'$ is reconstructed as  $x'_{\rm rec}= (x'_{1}+x'_{2})/2$,  
where $x'_{i}= \qprimesqrec / (W^2_{I,i}+\qprimesqrec)$ 
with
$W^2_{I,1}=(q'_{\rm rec}+ \xi  P)^2$  and $W^2_{I,2}= (\sum_{h} p_{h})^2$ 
where the sum runs over the HFS objects in the instanton band.
The  $W^2_{I,1}$ calculation is based on the scattered electron and the current jet, while
the $W^2_{I,2}$ reconstruction relies on the measurement of the hadronic final state objects 
in the instanton band.
The  $x'_{\rm rec}$ resolution achieved is about $50 \%$. 
As for the case of \qprimesqrecx, the reconstructed $x'_{rec}$ cannot be used to limit
the analysis to the kinematically allowed region $\xprimex \gsim\xprimeminx$.
However, $x'_{\rm rec}$ as well as  \qprimesqrec can be used to discriminate instanton processes
from the sDIS background.

Exploiting these observables, a multivariate discrimination technique is used 
to find the most sensitive set of observables to distinguish between signal and background~\cite{Sopicki:2013phd}.

\subsection{Comparison of Data to Standard QCD Predictions}
\label{sec:comparison}
\begin{figure}[htb]
    \epsfig{width=15.8cm,figure=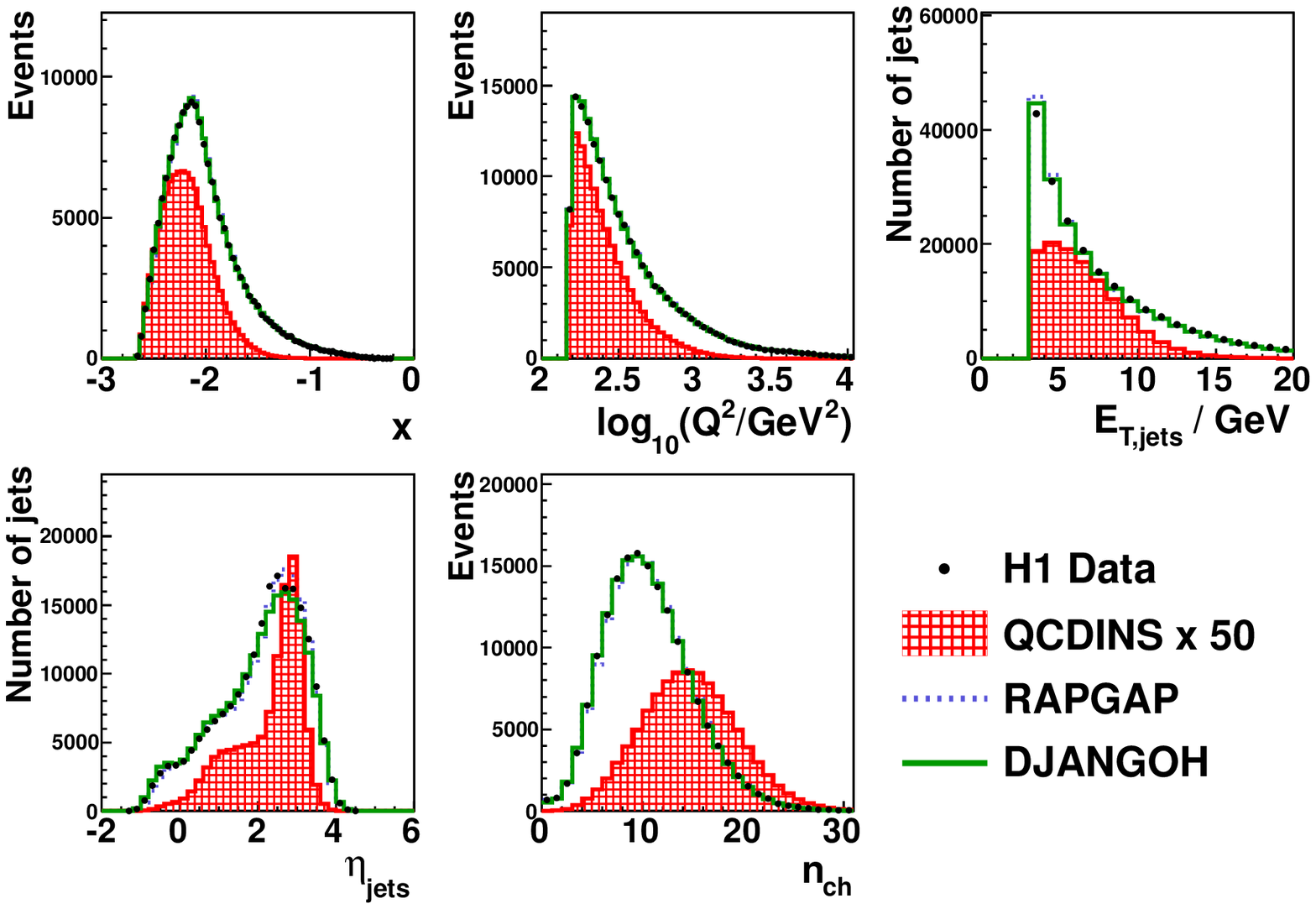  } 
 \vspace{-5.3cm} 
  \begin{picture}(50.,50.)
    \put(45. ,145.) {a)}
    \put(95. ,145.) {b)}
    \put(147.,145.){c)}
    \put(45.,94.) {d)}
    \put(95.,94.) {e)}
\end{picture}
  \caption{Distributions of (a) the Bjorken-scaling variable $x$,
          (b)  the photon virtuality $Q^2$,
	  (c)  the inclusive distribution of the transverse energy of the jets $E_{T, \rm jets}$, 
          (d) the pseudorapidity of the jets $\eta_{\rm jets}$
      and (e) the charged particle multiplicity $n_{\rm ch}$.  
   Data (filled circles), the RAPGAP and DJANGOH sDIS background predictions
   (dotted and solid lines) and the QCDINS signal
   prediction  scaled up by a factor of $50$ (hatched)  are shown.
  }
  
   \label{fig:CheckPlots}
\end{figure}

Both the RAPGAP and DJANGOH simulations provide a reasonable overall description of the experimental data 
in the inclusive DIS and  jet sample. To further improve the agreement between Monte Carlo
events and data, event weights are applied to match the jet multiplicities as a function of $Q^{2}$.  
The MC events are also weighted as a function of $P_{T}$ and $\eta$ of the most
forward jet in the Breit frame \cite{Roman,h1hiq2jets}.
Furthermore, the track multiplicity distribution is weighted.
The weights are obtained from the ratio of data to the reconstructed MC  distributions and
are applied to the events on the generator level.  
After these weights are applied, the simulations
provide a good description of the shapes and normalisation of the data distributions.
Examples of these control distributions are shown in figure~\ref{fig:CheckPlots}:
distributions of the kinematic variables $x$ and $Q^2$, 
the transverse energy of the jets 
$E_{T,\rm jets}$, the pseudorapidity of the jets $\eta_{\rm jets}$
in the hadronic centre-of-mass frame and the charged particle multiplicity $n_{\rm ch}$.

\begin{figure}[htb]
    \epsfig{width=15.8cm,figure=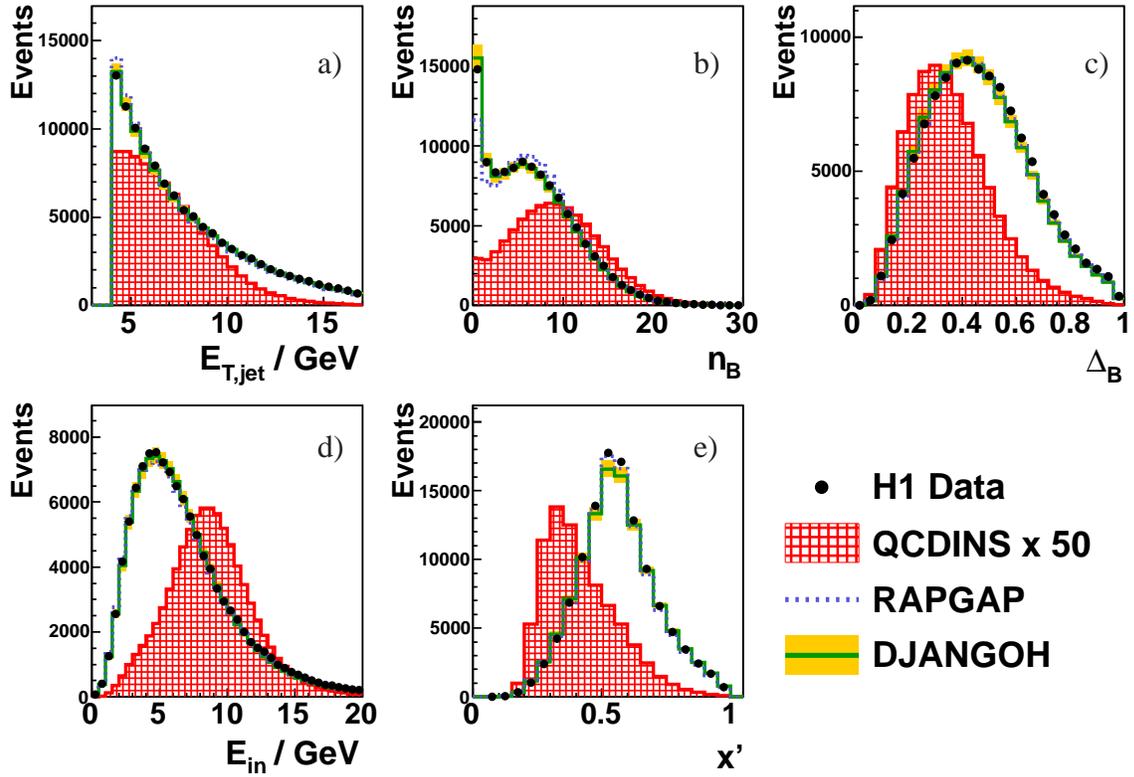  }     
 \vspace{-5.3cm} 
 \begin{picture}(50.,50.)
    \put(45. ,145.) {a)}
    \put(95. ,145.) {b)}
    \put(147.,145.){c)}
    \put(45.,94.) {d)}
    \put(95.,94.) {e)}
\end{picture}
  \caption{Distributions of the observables used in the multivariate analysis:
  (a) the transverse current jet energy \etjetx,
  (b) the charged particle multiplicity in the instanton band \nbx,
  (c) and (d) two variables measuring the azimuthal isotropy of the event, 
    \deltab and  $E_{\rm in}$, respectively, and
  (e) the reconstructed instanton kinematic variable $x'$.
   Data (filled circles), the RAPGAP and DJANGOH sDIS background predictions 
   (dotted and solid lines) and the QCDINS signal
   prediction  scaled up by a factor of $50$ (hatched),  are shown.   
   The error band, shown only for DJANGOH, represents the MC statistical and 
   systematic uncertainties added in quadrature.  
  }  
   \label{fig:ObsUsed}
\end{figure}

The measured distributions of the five  observables \etjetx, \nbx, $x'_{\rm rec}$, \deltab  and $E_{\rm in}$
are compared in figure~\ref{fig:ObsUsed} to the expectations from the standard DIS QCD models (RAPGAP, DJANGOH) and from 
the instanton model (QCDINS).
The data are reasonably well described
by the reweighted sDIS Monte Carlo simulations.  
The  models are able to describe the data
within $5-10\%$ except at very low and/or very large values of the given observable,
where  differences up to $ 20\%$ are observed.
The expected instanton distributions differ in shape from
the sDIS background. However, the magnitude of the expected signal is small and  advanced 
discrimination methods are required to enhance the signal to background ratio.

\section{Search for Instanton-Induced Events}
\label{sec:disc} 
A multivariate discrimination technique is  employed to increase the sensitivity to instanton processes.
The PDERS~({\it Probability Density Estimator with Range Search}) method  
as implemented in the TMVA ROOT package~\cite{TMVA2007} is used\footnote{
The PDERS method has been cross checked with other methods: the neural network 
MLP ({\it Multi-Layer Perceptron}) method and two variants of the decision tree method, BDT ({\it Boosted Decision Trees})
and BDTG ({\it Boosted Decision Trees with Gradient Boost})~\cite{Sopicki:2013phd}.}. 

The strategy to reduce the sDIS background 
is based on the observables \etjetx, \nbx, $x'$, \deltab  and $E_{\rm in}$.
This set of observables has been chosen since it provides the best 
signal to background separation~\cite{Sopicki:2013phd}.
Moreover, the distributions of these variables are overall well described by  both Monte Carlo simulations.
The distribution of the discriminator $D$ is shown in figure~\ref{fig:PDERSfullLinear}.
Taking into account the systematic uncertainties,
the  discriminator distribution  is described by  
the sDIS Monte Carlo simulations in the background dominated region.
For $D<0.2$  predictions and data agree  within systematic uncertainties.
The background events
are mainly concentrated at low discriminator values, while the instanton signal 
peaks at large values of the discriminator.  
At large $D$ both data and predicted background fall off steeply.

\begin{figure}[htb]
\centering  
    \epsfig{width=11.8cm,figure=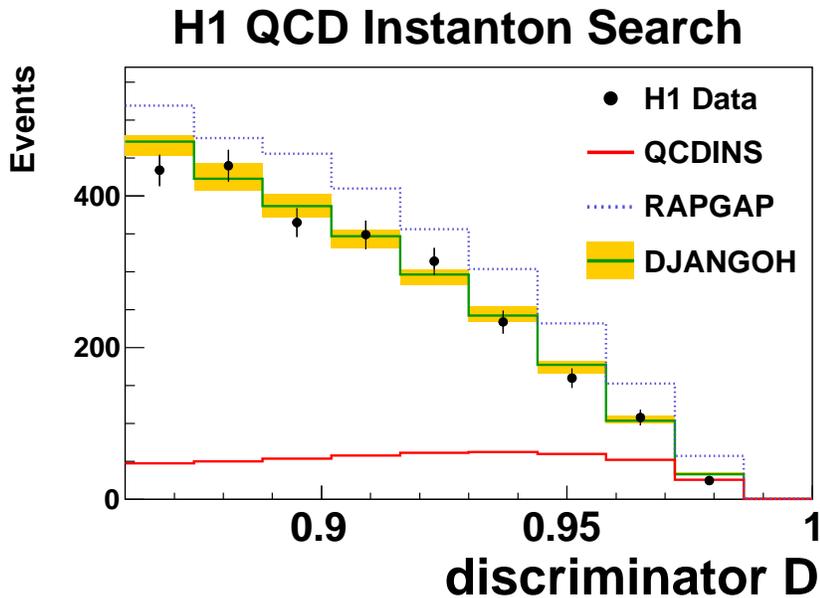  }        
  \caption{Distribution of the discriminator $D$ in the signal region $D > 0.86$.  
     Data (filled circles), the RAPGAP and DJANGOH sDIS background predictions
     (dotted and solid lines) and the QCDINS signal
     prediction (red line)  are shown.   
     The error band, shown only for DJANGOH, represents the MC statistical and 
     systematic uncertainties added in quadrature. 
   }
  \label{fig:PDERSzoom}  
\end{figure}

A signal region is defined for  $D > D_{cut}=0.86$, optimised for 
a determination of the instanton signal  from event counting. The distributions
of the expected instanton signal and of the background  are shown  in  
figure~\ref{fig:PDERSzoom}.
No excess of events is observed and
the DJANGOH MC describes
the data well, while the prediction of RAPGAP is systematically above the data.  
%

\begin{table}[h]
\begin{center}
\begin{tabular}{clccc}
\hline
         &     Data            &      DJANGOH         &          RAPGAP     &      QCDINS             \\
\hline
  $D>0.86$       &   $2430$            &  $2483^{+77}_{-90}$ &  $2966^{+~90}_{-103}$ & $473^{+10,+152}_{-12,-124}  $ \\
\hline  
\end{tabular}
\caption{Number of events observed in  data and expected from the DJANGOH
and RAPGAP simulations in the signal region. 
The quoted uncertainties include the experimental systematic uncertainties added in quadrature,
excluding  normalisation uncertainties.
For the  expected  number of instanton-induced events, the dominating uncertainty due to 
$\Lambda_{\overline{MS}}^{(3)}$ is also shown.
\label{tab:results}}
\end{center}
\end{table}

The expected and observed number of events are summarised in table~\ref{tab:results}.
In the signal region,  a total of  $2430$ events are observed in  data, while DJANGOH predicts
 $2483^{+77}_{-90}$  and RAPGAP $2966^{+~90}_{-103}$.
The uncertainties on the expected number of events include 
experimental systematic uncertainties and  small contributions from the
finite sample sizes. 
For the expected  number of instanton-induced events the dominating  uncertainty is due to 
$\Lambda_{\overline{MS}}^{(3)}$.

The following sources of  systematic uncertainties are propagated through the full analysis chain:
\begin{itemize}
 \item The energy scale of the HFS is known to a precision of $1\%$~\cite{h1hiq2jets,Roman}.
 \item Depending on the electron polar angle
       the energy of the scattered electron is measured with a precision of $0.5-1\%$~\cite{Aaron:2012qi}.
 \item The precision of the electron polar angle measurement  is $1$ mrad~\cite{Aaron:2012qi}.
 \item Depending on the electron polar angle, 
       the uncertainty on the electron identification efficiency ranges from  $0.5$ to $2\%$~\cite{Roman}. 
 \item The uncertainty associated with the track 
       reconstruction efficiency and 
       the effect of the nuclear interactions in  the detector material on the 
       efficiency of track reconstruction are estimated to be $0.5\%$ each~\cite{Brinkmann:2010zza}.
\end{itemize} 

The effect of these uncertainties on the expected signal and background  distributions
 is determined  by varying the corresponding
quantities by $\pm 1$ standard deviation in the MC samples and propagating  these variations through
the whole analysis.
The above systematic and statistical uncertainties added in quadrature
are shown in the figures and  in table~\ref{tab:results}. 
The included statistical  uncertainties due to 
the limited Monte Carlo statistics are approximately an order of magnitude smaller than
the experimental systematic uncertainties.

The main contributions to the experimental systematic uncertainties arise from  
the energy scale calibration of the scattered electron ranging from
$\sim4\%$ in the background dominated region to  $\sim1\%$ in the
signal region and from the energy scale of the HFS ranging from
$\sim1\%$ in the background region to $\sim2.5\%$ in the signal
region.
Uncertainties connected with 
the track reconstruction and
secondary interactions of the produced hadrons in the material
surrounding the interaction region
contribute to the systematic error 
in the signal region at a level of $\sim2\%$ each, and in the
background dominated region by less than $0.5\%$.
In the full range of the discriminator, the uncertainties on the electron identification and on the precision of
the electron polar angle are smaller than $0.5\%$ each.

Given the observed and expected  numbers of events, no evidence for QCD instanton-induced processes is observed.
In the following, the data are used to set exclusion limits.
\section{Exclusion Limits for Instanton-Induced Processes} 
\label{sec:limits}
\begin{figure}[htb]
\centering  
    \epsfig{width=11.8cm,figure=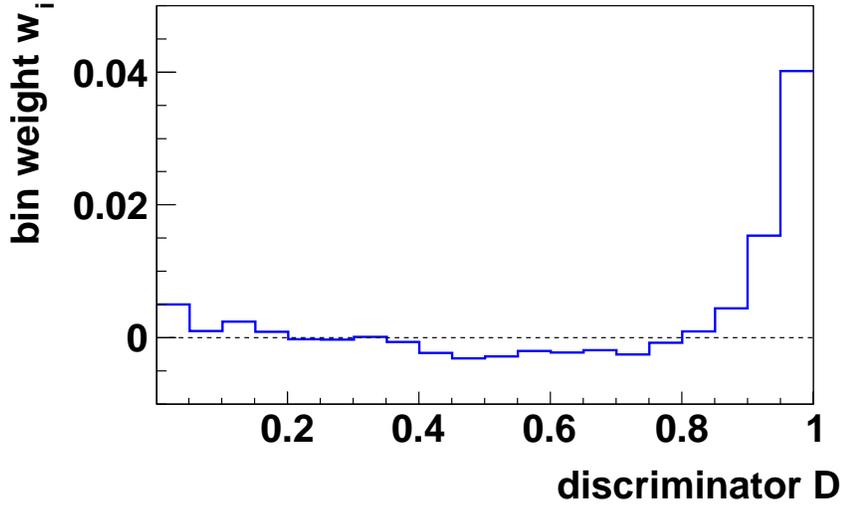 }        
  \caption{Distribution of the bin weights $w_i$ as a function of the discriminator $D$. The bin weights are calculated using the signal and background predictions together with their systematic uncertainties and the respective bin-to-bin correlations.
   }
  \label{fig:binweights}  
\end{figure}

The upper limit is determined from a CL$_{s}$ statistical analysis \cite{Read:2000ru,Junk:1999kv} 
using the method of fractional
event counting \cite{Bock:2004xz}.
A test statistic $X$ is constructed as a fractional event count of all events 
using the discriminator distribution:
\begin{equation} 
  X \,\, = \,\, \sum_{i=1}^{N_{\rm bin}} w_i n_i\,,
\label{eqn:X}
\end{equation}
where the sum runs over all bins, and $n_{i}$ is the number of events observed in bin $i$.
The weights $w_i$ are calculated from the predicted signal and background contributions and 
their  uncertainties, using an  appropriate set of linear equations~\cite{Bock:2004xz}. 
They are defined  in such a way as to ensure that only bins with both 
a large signal-to-background ratio  and small systematic 
uncertainties  enter with sizable weights into the test statistic $X$.
In case of negligible systematic uncertainties, the weights behave as  $w_i = s_i/(s_i + 2b_i)$ where
$s_i$ and  $b_i$ are the predicted number of signal and background events in a given bin $i$, 
respectively.
{\textchanged
In the presence of bin-to-bin correlated systematic uncertainties, the
weights may become negative in background-dominated regions. When
calculating the test statistics $X$ the negative weights correspond to
a subtraction of background contributions, estimated from data. The
distribution of the resulting weights $w_i$ is shown in
figure~\ref{fig:binweights}. Large positive weights are attributed to
bins in the signal region, $D>0.9$. Negative weights are assigned in
the region $0.4<D<0.75$.
}
A large number of MC experiments are generated by varying the expected number of events
in absence or presence of the signal within the statistical and systematic uncertainties. 
Systematic uncertainties are treated 
as Gaussian distributions and statistical  fluctuations are simulated using Poisson statistics.
If $1-CL_s>0.95$, the signal hypothesis is excluded 
at $95\%$ confidence level.

\begin{figure}[htb]

\centering 
 
    \epsfig{width=11.8cm,figure=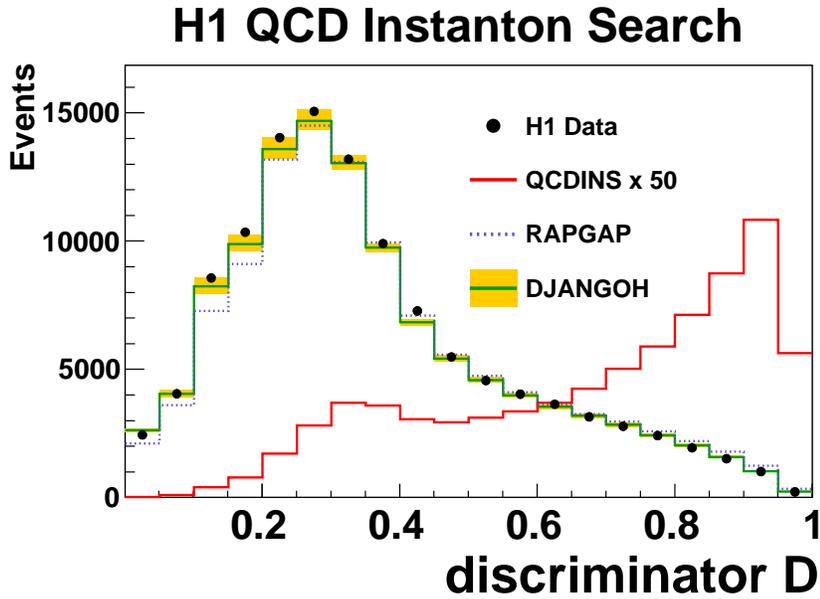  }

  \caption{Distribution of the discriminator $D$. 
   Data (filled circles), the RAPGAP and DJANGOH sDIS background predictions
   (dotted and solid lines) and the QCDINS signal
   prediction  scaled up by a factor of $50$ (red line)  are shown.   
   The error band, shown only for DJANGOH, represents the MC statistical and 
   systematic uncertainties added in quadrature. 
      }              
   \label{fig:PDERSfullLinear}  
\end{figure}

Limits are calculated using the full range of the discriminator distribution 
as shown in figure~\ref{fig:PDERSfullLinear}.
The following additional systematic uncertainties  are included 
in the exclusion limit calculation:
\begin{itemize}
 \item The normalisation uncertainty due to the precision of the integrated luminosity  measurement is $2.3\%$~\cite{compton-lumi}.
 \item The difference between the prediction from DJANGOH and RAPGAP  is assigned as  model uncertainty
       of the background estimation, 
       i.e. the difference between two background histograms in figure~\ref{fig:PDERSfullLinear}.
       This model uncertainty is large, $8-20\%$ and $13-46\%$, for small $D<0.2$ and large  $D>0.85$ 
       values of the discriminator, respectively.
       For  intermediate values of $D$  it amounts to $0.3-8\%$.
                 
 \item The uncertainty of the  background normalisation is  $1.1\%$.
        This uncertainty is estimated as
	$\epsilon = (N_{\rm Dj}-N_{\rm Rap})/N_{\rm Dj}$,
	where $N_{\rm Dj}$ and $N_{\rm Rap}$ are the total number of predicted events in the full discriminator
	range for the DJANGOH and RAPGAP MC simulations, respectively.	     
 \item The uncertainty of the predicted signal cross section  due to
       the uncertainty of $\Lambda_{\overline{MS}}^{(3)}$ (section \ref{sec:qcdtheo}) varies
      from 20 to 50\% depending on the region in $\qprime$ and $\xprime$. 
\end{itemize}

\begin{figure}[htb]
\centering
   \epsfig{width=11.8cm,figure=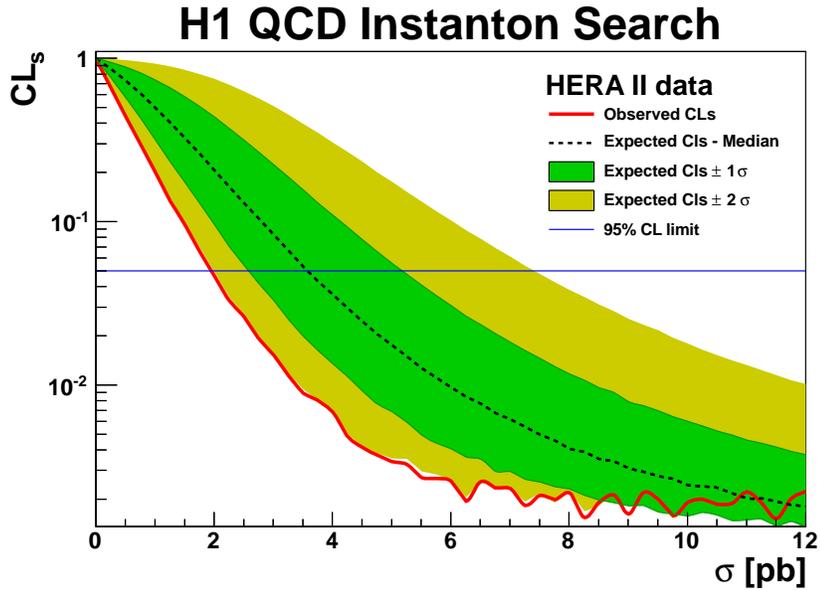  }	
   \caption{Observed CL$_{s}$ (solid line) as a function of the instanton cross section. 
             The $95\%$~CL limit is indicated by 
            a horizontal line. The dark and light bands correspond to  
	    $\pm1\sigma$ and $\pm2\sigma$ fluctuations of the expectation (dashed line).}
   \label{fig:CLSlimit} 	   
\end{figure}

Figure~\ref{fig:CLSlimit} shows the behaviour of the observed CL$_s$
as a function of the instanton signal cross section. 
In this study the total instanton cross section is taken as a free
parameter, whereas the signal shape is taken from the QCDINS simulation.
At $95\%$ CL, the observed limit is $2$~pb, as compared to
a median expected cross section limit of
$3.7^{+1.6}_{-1.1}(68\%)^{+3.8}_{-1.7}(95\%)$~pb. The first (second) set of
 uncertainties indicates the corresponding 
$\pm 1\sigma$ ($\pm 2\sigma$)  deviations
of the median expected cross section limit.
The  observed $-2\sigma$  deviation between the expected and observed limit
is caused by a downward fluctuation of the observed data test statistics $X$.
This downward fluctuation  receives contributions both from regions where 
the weights $w_i$ are positive and the data are below the background prediction and
from regions where the $w_i$ are negative and the data are somewhat larger than expected.

The QCD instanton model implemented in QCDINS,
restricted to the kinematic region defined by $\xprimeminx=0.35$ and
$\qprimesqminx=113$ \GeVsq{}, predicts a cross section of
$10\pm3$ pb, and thus is excluded by the H1 data.
Note that the  cross section uncertainty of $30\%$, stemming from
the variation of $\Lambda^{(3)}_{\overline {MS}}$, is already included
in the observed limit of $2$~pb.

\begin{figure}[htb]
\centering
   \epsfig{width=11.8cm,figure=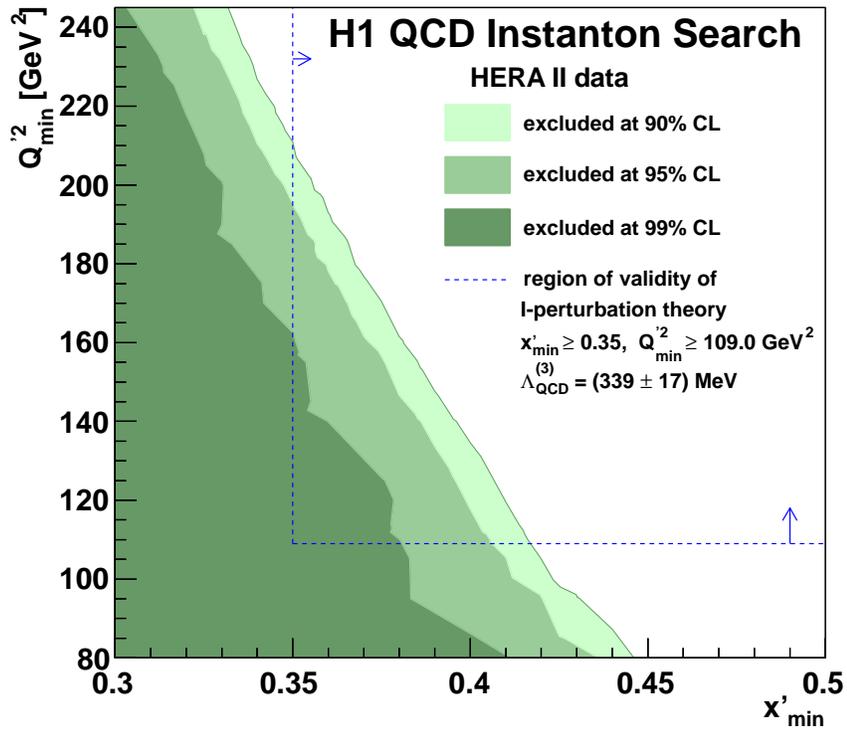  }	
   \caption{Instanton production exclusion limits as a function of 
      $\xprimeminx$ and
     $\qprimesqminx$. The regions excluded at  confidence levels of
     $90\%$, $95\%$ and $99\%$ are shown.  
     The   region of validity of instanton perturbation theory is indicated (dashed line).} 
   \label{fig:Cls339} 	   
\end{figure}

In order to assess the sensitivity  of the
instanton  cross section on the kinematic variables  $\xprimeminx$
and $\qprimesqminx$, limits are also determined as a function of the lower bounds 
$\xprimeminx$ and $\qprimesqminx$. As explained in section~\ref{sec:SimMC}, outside these  bounds 
the  instanton  cross section is set to zero.
The results are shown in figure \ref{fig:Cls339}, where the observed
confidence levels, using the QCDINS predictions, are shown in
the $(\xprimeminx,\qprimesqminx)$ plane.
At $95\%$ confidence level, parameter values
$\xprimeminx<0.404$ are excluded at fixed $\qprimesqminx=113$
\GeVsq{}. For fixed $\xprimeminx=0.35$, values of
$\qprimesqminx<195$ \GeVsq{} are excluded.
The exclusion regions depend somewhat on the choice of
$\Lambda^{(3)}_{\overline {MS}}$ and its uncertainty. In order to assess these effects,
the analysis was repeated for 
$\Lambda^{(3)}_{\overline {MS}}=340\pm 8$ MeV \cite{Agashe:2014kda} 
instead of $\Lambda^{(3)}_{\overline {MS}}=339\pm 17$ MeV . For this
choice, more stringent limits are obtained. For example, at fixed
$\qprimesqminx=113$ \gevsq the excluded range at $95\%$ confidence level
would change to $\xprimeminx<0.413$.

\begin{figure}[htb]
\centering
   \epsfig{width=11.8cm,figure=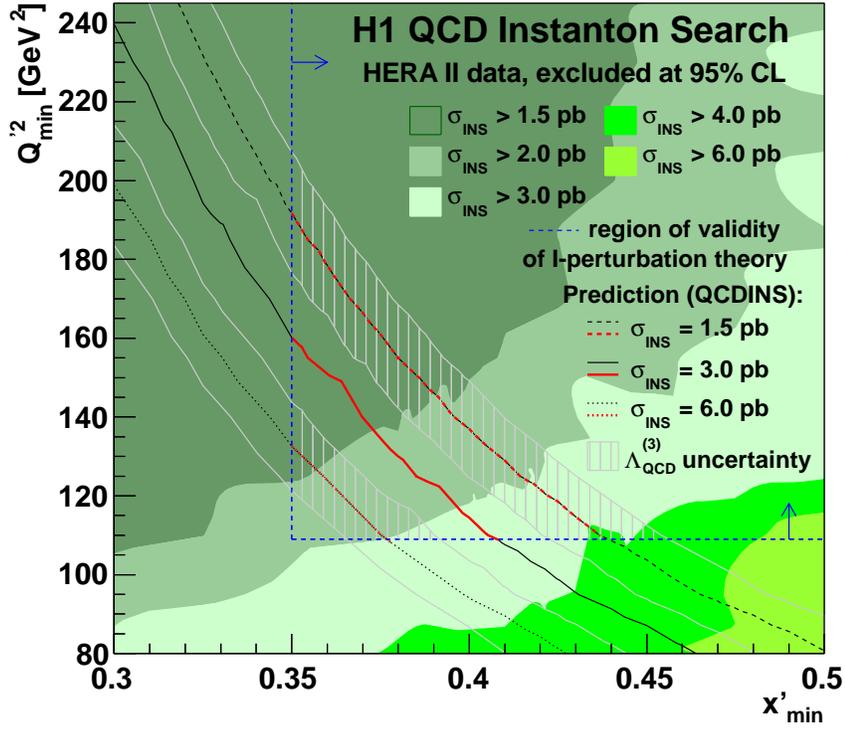  }	
   \caption{ Upper limits on the instanton  cross section
     at $95\%$~ confidence level, as a function of  $\xprimeminx$ and
     $\qprimesqminx$. Also shown are isolines of predicted fixed  instanton cross section
     and the effects of varying the QCD scale $\Lambda^{(3)}_{\rm QCD}$ defined in the $\overline{{\rm MS}}$ scheme  
     within  uncertainties.
     The  instanton cross section extrapolated beyond the  indicated region of validity of instanton 
     perturbation theory is shown as well.  } 
   \label{fig:LimitsPB} 	   
\end{figure}

A less model-dependent search is presented in figure
\ref{fig:LimitsPB}. Here, limits on the instanton cross section are determined
as a function of the parameters $\xprimeminx$ and
$\qprimesqminx$, using the signal shapes predicted by QCDINS.
No uncertainty on the instanton cross section
normalisation is included in this determination of the experimental
cross section limit.
The most stringent exclusion
limits of order $1.5$ pb are observed for large
$\qprimesqminx$ and small $\xprimeminx$. For increasing
$\xprimeminx$ the limits are getting weaker.
At the nominal QCDINS setting, $\xprimeminx=0.35$ and 
$\qprimesqminx=113$ \GeVsq, one expects to find back an exclusion
limit of $2$ pb, as discussed with figure \ref{fig:CLSlimit}. 
The limit in figure \ref{fig:LimitsPB}, however, is observed to be
somewhat better, because the theory uncertainty on the cross section
normalisation is included in figure \ref{fig:CLSlimit}
but not in figure \ref{fig:LimitsPB}.
\section{Conclusions}
A search for QCD instanton-induced processes is presented in 
neutral current deep-inelastic scattering
at the electron-proton collider HERA. The kinematic region is defined by 
the Bjorken-scaling variable $x > 10^{-3}$, 
the inelasticity $0.2< y < 0.7$ and the photon virtuality $150 < Q^2 < 15000$ GeV$^2$.
The search is performed using H1 data corresponding to 
an integrated luminosity of ~$351$ pb$^{-1}$.  

Several observables of the hadronic final state of the selected events are exploited 
to identify a potentially 
instanton-enriched sample. Two Monte Carlo models, RAPGAP and DJANGOH, are used 
to estimate the background from the standard NC DIS processes. 
The instanton-induced  processes are modelled by the program QCDINS. 
In order to extract the expected instanton signal a multivariate data analysis technique is used.
No evidence for QCD instanton-induced processes is observed.
In the kinematic region defined by  the theory cut-off parameters 
$\xprimeminx=0.35$ and $\qprimesqminx=113$ \GeVsq{}
an upper limit of $2$ pb  on the instanton cross section at $95\%$~CL is determined, as compared to
a median expected limit of $3.7^{+1.6}_{-1.1}(68\%)^{+3.8}_{-1.7}(95\%)$~pb. Thus, the
corresponding predicted instanton cross section of $10\pm 3$~pb is excluded by the H1 data. 
Limits are also set in the kinematic plane defined by \xprimeminx and \qprimesqminx.
These limits may be used to assess the compatibility of theoretical
assumptions such as the dilute gas approximation with H1 data, or to
test theoretical predictions of instanton properties such as their
size and distance distributions.

Upper cross section limits on instanton-induced processes  reported previously
by the H1 \cite{Adloff:2002ph} and ZEUS \cite{Chekanov:2003ww} collaborations
are above the theoretical predicted cross sections.
In a domain of phase space with a lower $Q^2$ range ($10 \lesssim  Q^2 < 100$ GeV$^2$),
H1 reported an  upper limit of $221$~pb at $95\%$~CL, about a factor five 
above the corresponding theoretical prediction. 
At high $Q^2$ ($Q^2 > 120$~GeV$^2$), the ZEUS Collaboration obtained an upper limit of
$26$~pb  at $95\%$~CL in comparison to a predicted cross section of $8.9$~pb.
In summary, compared to earlier publications, QCD instanton exclusion limits 
are improved by an order of magnitude
and  are challenging predictions
{\textchanged
based on perturbative instanton calculations with parameters derived from lattice QCD.
}

\section*{Acknowledgements}

We thank A. Ringwald and F. Schrempp for many helpful discussions
and for help with their computer program.
We are grateful to the HERA machine group whose outstanding
efforts have made this experiment possible. 
We thank the engineers and technicians for their work in constructing and
maintaining the H1 detector, our funding agencies for 
financial support, the
DESY technical staff for continual assistance
and the DESY directorate for support and for the
hospitality which they extend to the non DESY 
members of the collaboration.
We would like to give credit to all partners contributing to the EGI 
computing infrastructure for their support for the H1 Collaboration.


%
\clearpage

\clearpage

\clearpage



\end{document}